\documentclass[pra,12pt]{elsarticle}

\usepackage[a4paper, includefoot,
			left=1.5cm,
			right=1.5cm,
			top=2cm,
			bottom=2cm,			
			headsep=1cm,
			footskip=1cm]{geometry}
\usepackage{graphicx}
\usepackage{dcolumn}
\usepackage{bm}
\usepackage{color}
\usepackage[normalem]{ulem} 
\usepackage{amsmath}
\usepackage{amssymb}

\newcommand{\podd}{$\cal P$-odd~}
\newcommand{\todd}{$\cal T$-odd~}

\begin{document}
\title{Optical cycling in charged complexes with Ra--N bonds}

\author[1]{Timur Isaev}
\ead{timur.isaev.cacn@gmail.com}

\author[2,3]{Alexander V. Oleynichenko}
\ead{alexvoleynichenko@gmail.com}

\author[1,2]{Dmitrii Makinskii}
\ead{makinskii\_da@pnpi.nrcki.ru}

\author[1,2]{Andr\'{e}i Zaitsevskii}
\ead{zaitsevskii\_av@pnpi.nrcki.ru}

\affiliation[1]{ 
organization={Chemistry Dept., M. Lomonosov Moscow State University},
addressline={ Moscow 119991},
country={Russia} 
}%

\affiliation[2]{%
organization={B. P. Konstantinov Petersburg Nuclear Physics Institute of National Research Center ``Kurchatov Institute'' (NRC ``Kurchatov Institute'' – PNPI)}, 
addressline={Orlova Roscha, 1, 188300 Gatchina},
country={Russia}
}

\affiliation[3]{%
organization={Moscow Institute of Physics and Technologies (National Research University)},
addressline={Moscow region 141700},
country={Russia} 
}
\date{\today}

\begin{abstract}
The extension of laser cooling and trapping techniques to polyatomic molecular ions would have advanced scientific applications such as search of physics outside of the Standard Model, ultracold chemistry etc. We apply the Fock space relativistic coupled cluster method to study low-lying electronic states of molecular ions with Ra--N bonds, namely RaNCH$^+$, RaNH$^+_3$ and RaNCCH$^+_3$. Prospects of laser cooling of these species 
are estimated, and the peculiarities of unpaired-electron distributions are analyzed from the point of view of the molecular electronic structure. RaNH$^+_3$ and RaNCCH$^+_3$ are the first symmetric top molecular ions expected to be suitable 
for direct laser cooling. 
\end{abstract}

\maketitle

\section{Introduction}

The field of direct cooling of molecules with lasers has progressed immensely in the last decade. Starting with diatomic molecules \cite{DiRosa:2004, Barry:2014} the technique of molecular laser cooling was rapidly spread to polyatomic species \cite{Isaev:2016, Kozyryev:2017a, Mitra:2020}. One of the fields which is greatly influenced by developing molecular laser cooling is high-precision spectroscopy, where it is necessary to suppress stray electric and magnetic fields, which, in its turn, requires as much control over molecular degrees of freedom as possible. In the series of high-precision spectroscopic experiments on the ThO molecule and the HfF$^+$ molecular ion the strongest restrictions on the permanent electric dipole moment of an electron ($e$EDM), the property violating space ($\mathcal{P}$-odd) and time ($\mathcal{T}$-odd) parities~\cite{Landau:1957} have been recently set \cite{acme:2018, Roussy:2023}, and the next generation of experiments is expected to be performed with laser-coolable species~\cite{Isaev:2017, Kozyryev:2017b, Alarcon:22}. Nowadays molecules and molecular ions containing radioactive heavy nuclei are recognized as perspective objects for searching for the $\mathcal{P}$-odd and $\mathcal{T}$-odd effects enhanced in  such systems~\cite{Isaev:10a, GarciaRuiz:2020, Oleynichenko:2022, Chamorro:22}. Rather common principles of the design of laser-coolable polyatomic molecules were formulated in ~\cite{Isaev:2016} (the special case of molecular ions is considered in ~\cite{Ivanov:2020b,Wojcik:2023}) and recently we proposed several charged laser-coolable polyatomic Ra-containing molecules that are expected to be prospective for searches of new \podd and \todd fundamental forces \cite{Isaev:2022a}.
Radium compounds seem to be among the most promising candidates for such searches not only due to quite large enhancement factors of $\mathcal{T,P}$-odd effects (typical for all compounds containing heavy nuclei), but mainly due to a large variety of isotopes available for spectroscopic experimental studies and possessing nuclear properties valuable for different types of experiments. For example, the relatively long-lived $^{224}$Ra (half-life 3.6~d), $^{226}$Ra (1600~y), $^{228}$Ra (5.8~y) isotopes studied in recent papers on the RaF molecule spectroscopy~\cite{Udrescu:2021,Athanasakis:RaF:2023} have zero nuclear spin and thus are perfect for experiments aimed at {the} $e$EDM searches, whereas nuclei of other isotopes like $^{223}$Ra (half-life 11.4~d) and $^{225}$Ra (1.49~d) are predicted to have strongly enhanced $\mathcal{P},\mathcal{T}$-violating nuclear Schiff and magnetic quadrupole moments (MQM)~\cite{Dobaczewski:05,Flambaum:22}, thus allowing one to probe such effects in experiments with molecules containing these isotopes. Note that the open-shell molecules discussed in the present work suit well for experiments aimed at the $e$EDM and nuclear MQM searches.

The basic principles of molecular laser-cooling and related experimental techniques were reviewed in details in~\cite{Tarbutt:2015, Augenbraun:2023}. In brief, molecules amenable for direct laser cooling typically have a restricted domain referred to as an optical cycling center (OCC)~\cite{Li:2019,Ivanov:2019} where the main changes in electronic wavefunction occur under the interaction with light. Crucial property of the OCC is that electronic transitions associated with such an optical cycling center involve non-bonding molecular orbitals 
(molecular spinors)
\cite{Isaev:2016}, initially described and classified for diatomic laser-coolable molecules~\cite{Isaev:2010}. In this case several lowest vibronic transitions are characterised by sufficiently diagonal Franck-Condon (FC) matrix, and a cooling scheme with one main cooling laser and a few repumping lasers, gathering a population from {lower-state} excited vibrational levels back to the ground vibrational level of the {upper} electronic state, is feasible~\cite{DiRosa:2004}. Typically to estimate the degree of a cooling loop closure the sum of the largest FC factors for 
transitions from the ground vibrational level of the excited electronic state to a few vibrational levels of the ground electronic state is used. The rate of decrease of FC factors with the increase of the lower-state vibrational quantum number depends heavily on small variation of electronic wavefunctions in the vicinity of an optical cycling center and is thus sensitive to chemical environment (relatively) close to an OCC.
In principle, it allows one to tune the optical properties of OCC
by a careful choice of ligands bound to an OCC (see e.~g. \cite{Isaev:2018c, Dickerson:2021}). 
The most economical and effective way to make a proper choice implies electronic structure simulations employing state-of-the art \textit{ab initio} techniques. Several studies have been performed to find some general route to enforce the desired behavior of the Franck-Condon factors for vibronic transitions between working electronic levels by the thorough selection of ligands~\cite{Ivanov:2019, Zhu:2022, Dickerson:2021}.

A large amount of previous theoretical and experimental papers on laser-coolable polyatomics considered organometallic molecules of the type M-O-R, where M stands for a Group II metal atom, R is an organic radical and the oxygen atom serves as a bridge effectively screening the OCC located on the M atom from the radical \cite{Kozyryev:2016b}. The most notable systems of this type are CaOCH$_3$~\cite{Mitra:2020}, RaOCH$_3$~\cite{Zakharova:22},  YbOCH$_3$~\cite{Augenbraun:YbOCH3:21} etc. However, oxygen is not the only atom which can be used as a bridge screening an OCC from ligand, it could be also N or C~\cite{Ivanov:2019,Ivanov:2020}.
That would be also interesting to consider some other 
compounds 
with an optical cycling center 
located on a heavy atom (like Ra) 
and a nitrogen ``bridge'' and to analyse how optical properties of an OCC in compounds of the Ra-N-R type depend on 
the choice of
R. From this point of view it seems promising to examine 
systems
with the N$\equiv$C triple bond near to the OCC, since laser coolability was recently predicted for the MCCH-type molecules 
with triple C$\equiv$C bond
(MgCCH, CaCCH, SrCCH~\cite{Xia:CaCCH:21,Changala:23}).
The simplest Ra-containing system of this type is the RaNCH$^+$ ion. However, the most promising objects for parity violation search experiments are symmetric top molecules due to the presence of $K$-doublets, thus it is reasonable to consider also the RaNCCH$_3^+$ ion. The third ion considered in the present work is RaNH$_3^+$, also possessing a symmetric top geometry. It is 
isovalent
to the neutral CaCH$_3$, MgCH$_3$~\cite{Isaev:2016}, BaCH$_3$ and YbCH$_3$~\cite{Chamorro:22} molecules previously shown to be promising for laser cooling. It is worth noting that the positive charge of the structures considered seems to be highly desirable from the point of view of possible future experiments with such systems since it can greatly facilitate the trapping and manipulation of such molecules. To the best of the authors' knowledge, molecular ions have not been laser-cooled yet (for translational degrees of freedom).

In the present work we analyze 
electronic structure 
and low-lying excited states of 
RaNCH$^+$, RaNH$_3^+$ and RaNCCH$_3^+$ 
molecular ions 
which were shown to have rather strong Ra--N bond~\cite{Isaev:2022a} in their ground states. The interrelations between electronic density distribution of natural transition spinors, molecular geometry in excited states, and laser coolability are revealed, paving the way towards the rational design of the N-bridge organometallics with highly closed optical loops. 

\section{Computational scheme}

Calculations of low-lying electronic states of RaX$^+$, X=NCH, NH$_3$, NCCH$_3$, were performed within the Fock-space relativistic coupled cluster (FS RCC) method with the cluster operator expansion restricted to single and double excitations (CCSD)~\cite{Visscher:01,Eliav:Review:22}. The computational scheme closely resembled that used in Refs.~\cite{Isaev:2021a,Isaev:2022a}. An important difference is the use of non-local generalized relativistic pseudopotentials, GRPPs~\cite{Mosyagin:06amin} (rather than their semilocal components employed in ~\cite{Isaev:2022a}) to simulate 60 core electrons of Ra and relativistic effects on C and N atoms; the use of full GRPPs seems to be essential for more accurate reproduction of excitation energies~\cite{Oleynichenko:23libgrpp}. The triple zeta quality all-electron basis sets for light elements augmented with diffuse functions for C and N were exactly the same as in~\cite{Isaev:2022a} whereas a new generally contracted basis 
$(17s\,14p\,12d\,11f\,7g\,6h)$ / $[7s\,8p\,8d\,6f\,4g\,3h]$ compatible with GRPP was constructed for Ra (see Supplementary materials). The Fock space scheme corresponded to the use of the RaX$^{2+}$ ground state solution of the spin-orbit-coupled Hartree-Fock equations as a Fermi vacuum, whereas the target RaX$^+$ states are described within the $0h1p$ (one particle outside of the closed shell) Fock space sector. The $0h1p$ model space was spanned by Slater determinants with a single electron on one of 18 lowest virtual ``active'' spinors of RaX$^{2+}$. During the calculations of model natural spinors (i.~e. natural spinors of model-space parts of electronic wavefunctions, see ~\cite{Isaev:2022a}) the active set additionally included all other virtual spinors with negative orbital energies; the corresponding Slater determinants were treated as ``buffer'' ones~\cite{Zaitsevskii:23}.

Most calculations were carried out with the frozen ligand geometries. The smallness of ligand deformations upon the attachment of the ground-state Ra$^+$ ion has been demonstrated in~\cite{Isaev:2022a}. To 
check the validity of this approximation
in optical cycling studies, we performed 
full geometry optimization at the FS RCC level for the ground and first excited states of RaNCH$^{+}$ and RaNH$_3^{+}$, using slightly reduced basis sets ($[6s\,6p\,6d\,4f\,2g]$ for Ra, $[5s\,4p\,3d\,1f]$ for N, C and $[3s\,2p]$ for H); the results clearly indicate that the changes of ligand geometries upon low-energy electronic excitations are negligible (see Table~\ref{bending}). The values of {equilibrium} internuclear separations and bond angles in ligands for calculations employing enlarged basis sets (see Supplementary materials) were obtained by adding the (rather small!) changes upon the attachments of Ra$^+$ ($^{\,2}S_{1/2}$) estimated at the relativistic DFT level~\cite{Isaev:2022a} to the experimental equilibrium parameters of the free ligands taken from Refs.~\cite{doi-10.1139-p57-134, BOTSCHWINA1997119, Fujimoto2015}.

In order to partially reduce the effect of the neglect of higher cluster amplitudes on the shapes of resulting potential energy cross sections, we evaluated excited-state energies by adding the corresponding geometry-dependent FS RCC excitation energies to the ground-state potential calculated within the single-reference open-shell relativistic coupled cluster approximation accounting perturbatively the contribution from connected triple excitations, RCCSD(T) (cf. Refs~\cite{Isaev:2021a,Isaev:2022a}). We shall refer to the resulting estimates as 
FS RCCSD/ RCCSD(T).

Radiative lifetimes of the lowest excited states were estimated {\it via} evaluating transition dipole moments according to the approximate scheme described in ~\cite{Zaitsevskii:23tho}. This scheme implies the retention of linear and quadratic terms of the expansion of 
the 
FS RCC effective dipole moment operator in powers of cluster amplitudes and provides significant improvements over the simple model-space approximation employed in our previous study~\cite{Isaev:2022a}.

The solutions of the relativistic Hartree-Fock problem and molecular integrals transformed to the molecular spinor basis as well as RCCSD(T) ground-state energies were obtained using the DIRAC19 program~\cite{DIRAC:19,DIRAC:2020} augmented with the LIBGRPP toolkit~\cite{Oleynichenko:23libgrpp} for evaluating molecular integrals with GRPPs; the EXP-T code~\cite{EXPT:20} was employed for FS RCC calculations of both excitation energies and transition dipole moments.

In order to estimate prospects to build quasi-closed optical cycles involving the first excited and ground states of the molecular ions under study, we evaluated the FC factors for the transitions $(v^{\prime}_{\parallel}=0,\;v^{\prime}_{\perp}=0)$ $\to$ $(v^{\prime\prime}_{\parallel}=0,\,1,\,2, \;v^{\prime\prime}_{\perp}=0)$, where $v_{\parallel}$ and $v_{\perp}$ denote the vibrational quantum numbers for the Ra--N stretching and bending modes, respectively; single and double primes correspond to the upper and lower electronic states. The vibrational problems were solved within the rigid-ligand approximation. The coupling of stretching and bending modes as well as the effect of rotations were neglected. Stretching vibrational functions for {the} three lowest levels were obtained by solving numerically the one-dimensional Schr\"{o}dinger equations~\cite{Sundholm}. Since only the lowest bending vibrational levels were considered in the present study, the harmonic approximation appears {to be} sufficiently accurate in this case. Within the chosen approximation, each FC factor is represented by a product of single-mode factors. 

\begin{figure}[h]
    \centering
    \includegraphics[width=0.9\textwidth]{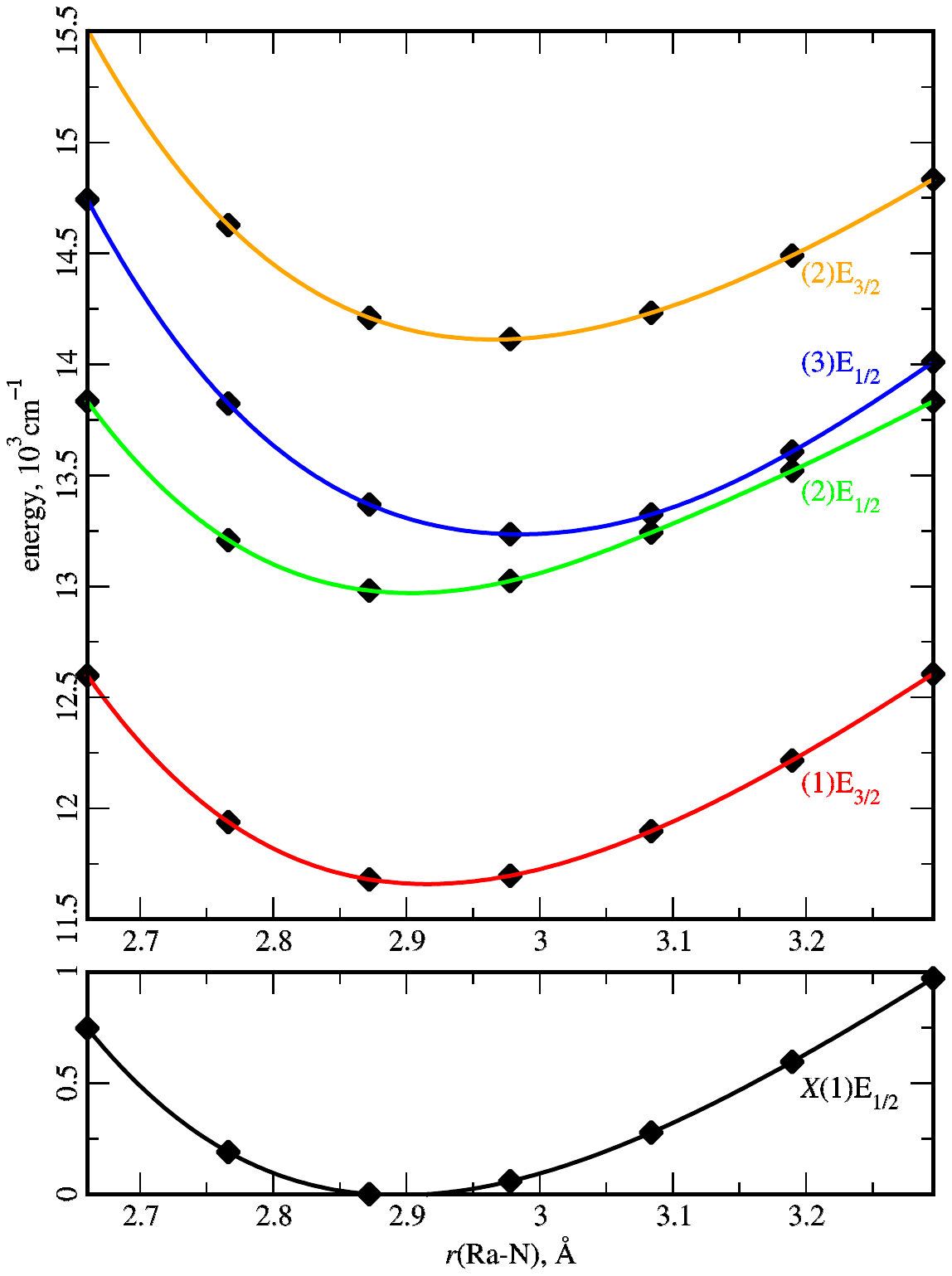}
    \caption{FS RCCSD / RCCSD(T) energies of 
low-lying electronic states of symmetric-top ($C_{3v}$) RaNH$_3^+$ complex as functions of of the Ra--N internuclear separation.}
    \label{fig:ranh3curves}
\end{figure}

\begin{figure}[h]
    \centering
    \includegraphics[width=0.9\textwidth]{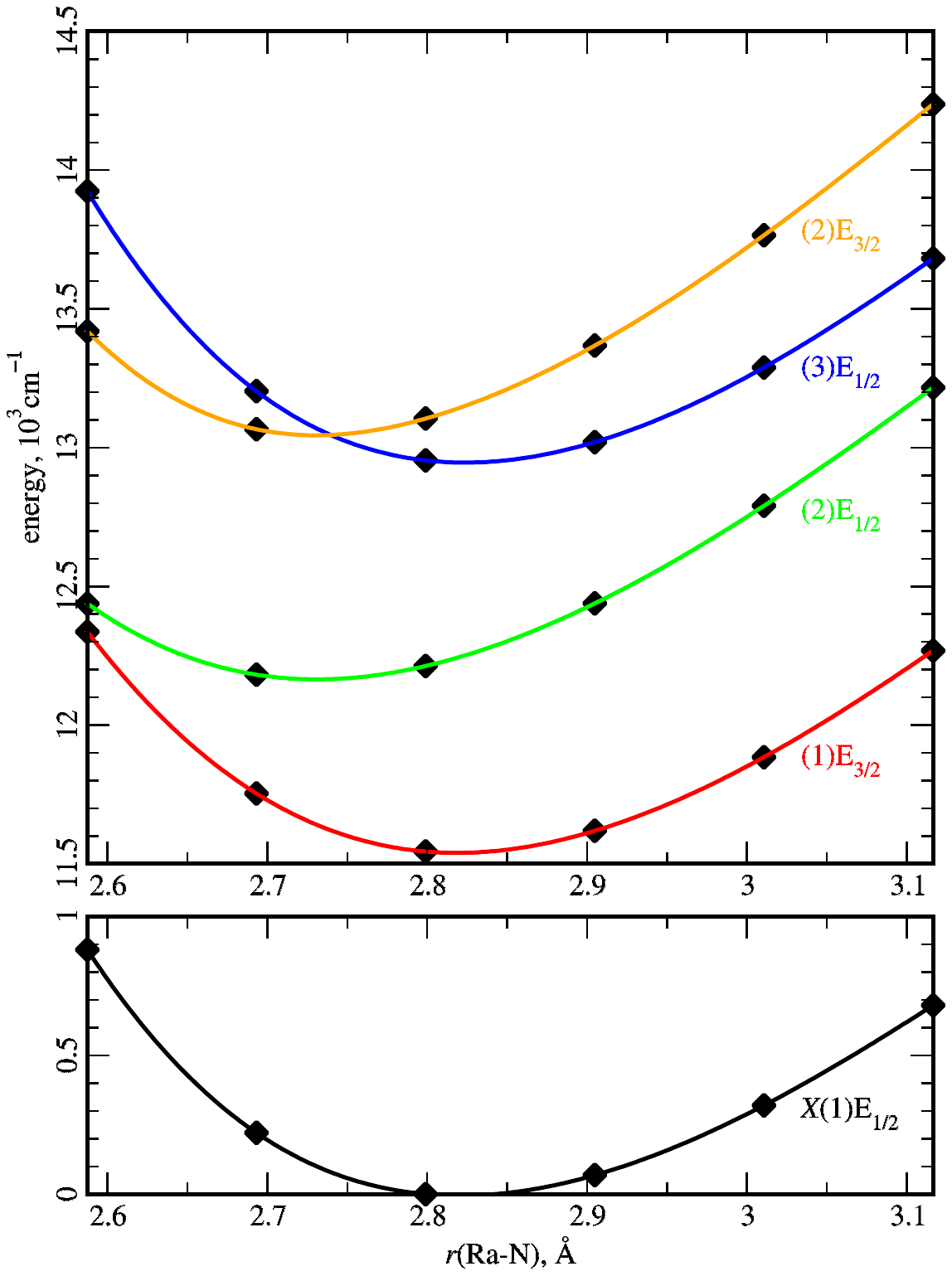}
    \caption{FS RCCSD / RCCSD(T) energies of 
low-lying electronic states of a symmetric-top ($C_{3v}$)RaNCCH$_3^+$ complex as functions of the Ra--N internuclear separation.}
    \label{fig:rancch3curves}
\end{figure}

\begin{table}
\caption{\label{optgeometry} Changes $\Delta$ of FS RCC equilibrium geometry parameters upon lowest-energy electronic excitation for RaNCH$^+$ and RaNH$_3^+$.}
\renewcommand{\arraystretch}{1.3}
\begin{tabular*}{\columnwidth}{l@{\extracolsep{\fill}}lcc}
\hline
\hline
Complex & Parameter & Ground state &  $\Delta$ \\
\hline
RaNCH$^+$ & $r$(Ra--N), \AA & 2.933 & $-$0.055\\
          & $r$(N--C),  \AA & 1.151 &   0.001\\
          & $r$(C--H),  \AA & 1.072 &  $-$0.001\\[1ex]
RaNH$_3^+$& $r$(Ra--N), \AA & 2.932 &   0.023\\
          & $r$(N--H),  \AA & 1.019 &  $-$0.001\\
          & $\angle$ RaNH, $^{\circ}$  &    113.6  & 0.0\\
\hline
\hline
\end{tabular*}
\end{table}

\begin{table*}
\caption{\label{bending} {Equilibrium Ra--N internuclear separations ($r_e$), adiabatic term energies ($T_e$), harmonic vibrational constants for stretching and bending modes ($\omega_\parallel$ and $\omega_{\perp}$, respectively), ground-state dissociation energies ($D_e$), and radiative lifetimes of excited states ($\tau$) for cationic complexes of Ra.}}
\renewcommand{\arraystretch}{1.3}
\begin{tabular*}{\textwidth}{l@{\extracolsep{\fill}}ccccccc}
\hline
\hline
 Complex & State & $r_e$, \AA & $T_e$, 10$^3$ cm$^{-1}$ & $\omega_\parallel$, cm$^{-1}$ & $\omega_{\perp}$, cm$^{-1}$ &$D_e$, 10$^3$ cm$^{-1}$ & $\tau$, $\mu$s \\
\hline
RaNCH$^+$ & $X(1)1/2$ & 2.897 & 0 & 177 & 124 & 6.99\\
 & (1)3/2 & 2.850 & 11.28 & 156 & 135 && 21\\[1ex]
RaNH$_3^+$ & $X(1)E_{1/2}$ & 2.893 & 0 & 202 & 424 & 7.84\\
  & $(1)E_{3/2}$ & 2.916 & 11.66 & 204 & 405 && 61\\[1ex]
RaNCCH$_3^+$  & $X(1)E_{1/2}$ & 2.821 & 0 & 143 & 71 &9.21 \\
 & $(1)E_{3/2}$ & 2.818 & 11.54 & 146 & 75 && 13\\
\hline
\hline
\end{tabular*}
\end{table*}

\begin{figure}[h]
    \centering
    \includegraphics[width=0.45\textwidth]{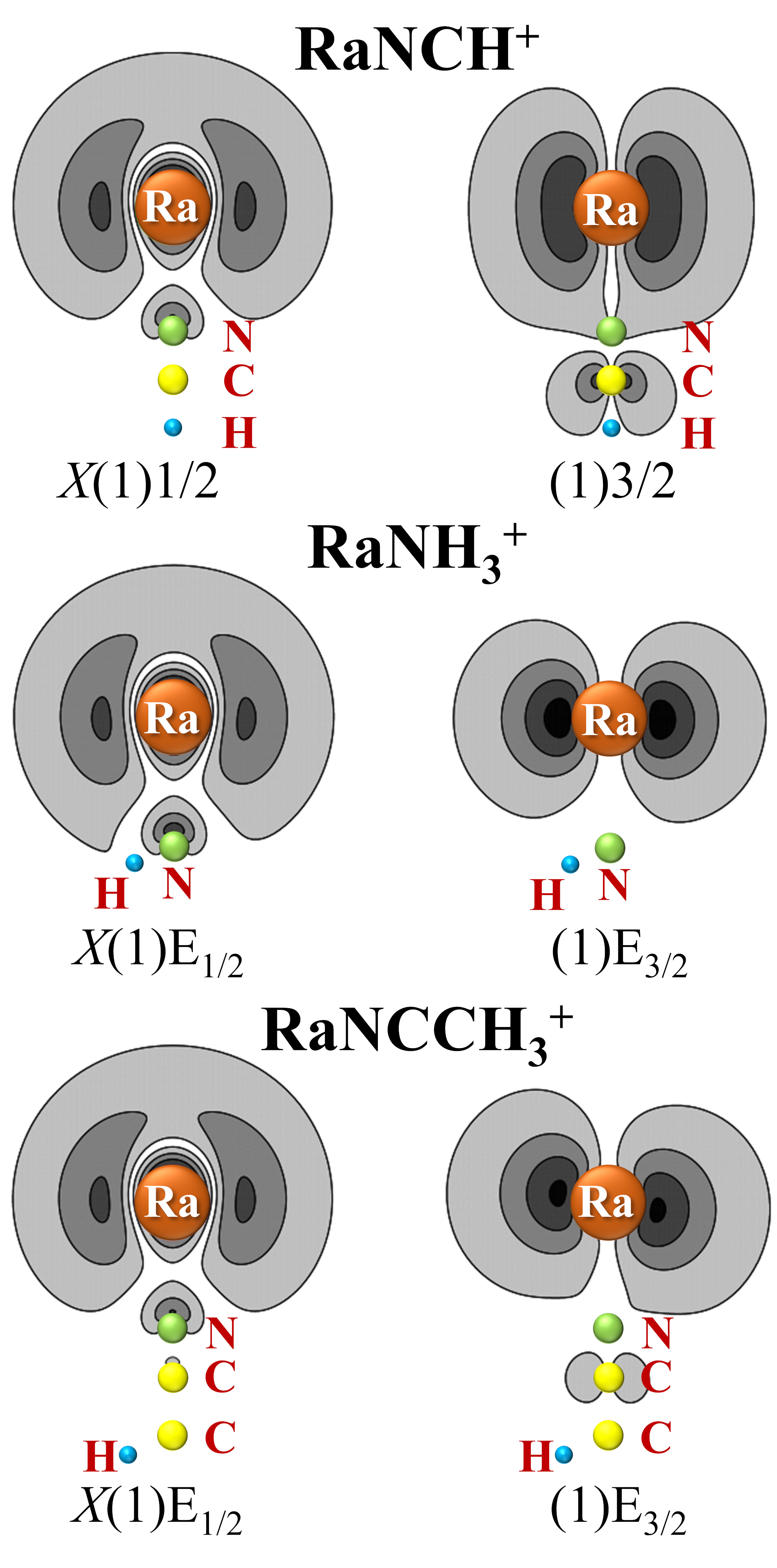}
    \caption{
Plots of absolute values of model natural 
spinors for the 
ground and first excited states of RaNCCH$_3^+$, RaNH$_3^+$ and RaNCH$^+$ in the planes passing through Ra, N, and one H nuclei. }
    \label{fig:ntsvis}
\end{figure}

\begin{table}[htp]
\caption{Cumulative values of Franck-Condon factors 
for $(0,0,0)\to(v^{\prime\prime},0,0)$ vibronic transitions 
from the first excited 
to the ground electronic state of RaX$^{+}$ complexes}\label{fcf}
\begin{center}
\renewcommand{\arraystretch}{1.3}
\begin{tabular*}{\columnwidth}{l@{\extracolsep{\fill}}ccc}
\hline
\hline
& RaNCH$^+$ & RaNH$_3^+$  & RaNCCH$_3^+$  \\
\hline
$v^{\prime\prime}=0$     & 0.8523 & 0.9787 & 0.9986 \\
$v^{\prime\prime}=0,1$   & 0.9626 & 0.9994 & 0.9991 \\
$v^{\prime\prime}=0,1,2$ & 0.9815 & 0.9995 & 0.9991 \\
\hline
\hline
\end{tabular*}
\end{center}
\label{default}
\end{table}%

\section{Results and Discussion}

Equilibrium geometries for all electronic states of the complexes under study arising from ground-state ligands and 
$7s$, $7p$ and $6d$ states of the
Ra$^+$ atomic ion are highly symmetric ($C_{\infty v}$ for RaNCH$^+$, $C_{3 v}$ for RaNH$_3$ and RaNCCH$_3$). The cross sections of potential energy surfaces along the Ra--N internuclear separation are presented in Figs.~\ref{fig:ranh3curves} and~\ref{fig:rancch3curves} (see also Fig.~1 in ~\cite{Isaev:2022a} for the potential energy curves of the RaNCH$^+$ ion); Table~\ref{bending} provides the FS RCCSD / RCCSD(T) molecular constants for the ground and lowest excited states evaluated within the rigid-ligand approximation. 

Firstly one can see that the approximation  applied in the {present} study  {to}  RaNCH$^+$\cite{Isaev:2022a} (the neglect of contributions to FC-factors from the bending modes) could not affect 
the conclusion concerning the prospects of laser cooling, as the sum of FC factors for vibronic transitions $(v^{\prime}_{\parallel}=0,\;v^{\prime}_{\perp}=0)$ $\to$ $(v^{\prime\prime}_{\parallel}=0,\,1,\,2, \;v^{\prime\prime}_{\perp}=0)$ is also close to unity. 

The sum of three largest Franck-Condon 
factors ( {recall} that this number is defined by a technical ability of using such or another number of separate lasers in a laser cooling experiment) gradually decreases in the series RaNH$_3^{+}\gtrapprox$ RaNCCH$_3^{+}>$ RaNCH$^+$. 

The main features of the model natural spinors which, in the particular case of the $0h1p$ Fock space sector, describe the unpaired electron distributions and coincide with the model natural transition spinors for the excitations between the states described within this sector~\cite{Isaev:2021a}, are in agreement with the trends mentioned above. In the ground states of all the three complexes, the unpaired electron occupies the non-bonding spinor localized on Ra. A strict localization of the unpaired electron on the radium ion is also observed for the excited $(1){\rm E}_{3/2}$ state of RaNH$_3^+$ and with some reservations, for the similar state of RaNCCH$_3^+$. 

The Ra--N bond strength and length can change non-negligibly upon electronic excitation due to a rather slight mixing of $d\pi$ functions of Ra$^+$ with virtual (mainly antibonding)  $\pi$-like orbitals localized on the N--C bond. This effect leads to a relatively slow decrease in $( v^{\prime}_\parallel=0) \to (v^{\prime\prime}_\parallel)$ FC factors with increasing $v^{\prime\prime}_\parallel$, as occurs in the case of RaNCH$^+$. However, one can suppress this effect by an appropriate choice of the substituent at the carbon atom (RaNCCH$^+_3$) or simply by using N-containing ligands without multiple N--C bonds (RaNH$^+_3$).
Thus, molecules with Ra-N bonds can be also promising candidates for direct cooling with lasers, like
M-O-R (Metal-Oxygen-Radical), 
M-S-R etc. compounds considered earlier \cite{Mitra:2020, Isaev:2018c}. 

Radiative lifetimes of the lowest-lying excited electronic states presented in Table~\ref{bending} are nearly three orders of magnitude larger than those of the typical neutral molecules which were successfully laser-cooled to date due to the nearly pure $7s-6d$ nature of this electronic transition. Actually the same situation was previously predicted for the AcOH$^+$ ion~\cite{Oleynichenko:2022}. Despite the fact that the short excited state lifetime is of vital importance for laser cooling of neutral molecules, it seems 
not compulsory for molecular ions, since they can be trapped relatively easily and then laser-cooled to the Doppler limit, approximately corresponding to several nK for the transitions considered. For example, such a radio frequency trap was used in the series of $e$EDM experiments on the HfF$^+$ ion (see~\cite{Roussy:2023} and references therein).

\section{Conclusion}

In the present work the laser coolability of several molecular ions of the 
Ra-N-R type, where 
Ra plays the role of 
optical circulation center and R denotes an atom or radical, was estimated via high-precision \textit{ab initio} relativistic calculations of potential energy surfaces, thus continuing the previous study from ~\cite{Isaev:2022a}. The analysis of spatial distributions of model natural spinors for the systems considered allows one to reveal the interconnections between the properties of the Franck-Condon vector desired for laser coolability and the presence of 
a certain degree of $\pi$-type bonding between an optical cycling center and a neighbouring multiple chemical bond in an excited state of a molecule. 
This results in a shortening of a chemical bond
upon excitation, which 
leads to 
a significant population leakage rate. 
One can expect that a simple analysis of natural transition spinors can be useful in at least rejection of molecules hopeless for direct laser cooling due to the break of parallelity of potential energy surfaces arising from additional bonding in excited states.

Two molecular ions with the Ra--N bond, namely RaNH$_3^+$ and RaNCCH$_3^+$, have been demonstrated to be promising systems for direct laser cooling. It should be underlined that chiral isotopologues of both ions can in principle be produced for sensitive experimental probes for the direct detection of $\mathcal{P}$-odd interactions (see~\cite{Isaev:2018c} and references therein). In this respect accurate calculations of constants characterizing different types of $\mathcal{P}$- and $\mathcal{T,P}$-odd interactions in these two molecular ions seem to be desirable, and such calculations may be the goal of future research.

\section{Credit authorship contribution statement}

\textbf{Timur Isaev}: Conceptualization, Methodology, Formal analysis, Writing – original draft, Writing – review \& editing. \textbf{Alexander Oleynichenko}: Software, Formal analysis, Writing – review \& editing. \textbf{Dmitrii Makinskii}: Software, Formal analysis. \textbf{Andrei Zaitsevskii}: Conceptualization, Methodology, Investigation, Validation, Writing – original draft, Writing – review \& editing.

\section{Declaration of competing interest}

The authors declare no competing interest.

\section{Data availability}

The data that support the findings of the study are available within the article and its Supporting information, which includes 
information on the basis set for Ra employed in the present work, frozen geometry parameters of ligands and potential surface cross sections of ground and excited states of the RaNH$_3^+$ and RaNCCH$_3^+$ molecular ions. The other data are available on request.

\section{Acknowledgments}

Authors are grateful to Anastasia~V. Bochenkova and Robert Berger for useful discussions. The calculations have been carried out using computing resources of the federal collective usage center Complex for Simulation and Data Processing for Mega-science Facilities at National Research Centre ``Kurchatov Institute'', http://ckp.nrcki.ru/and using the equipment of the shared research facilities of HPC computing resources at Lomonosov Moscow State University. Relativistic electronic structure modeling of molecular ions performed by TI, DM and AZ at the M. V. Lomonosov Moscow state university was supported by the Russian Science Foundation (grant No. 21-42-04411).

\bibliographystyle{apsrev}


\end{document}